\documentclass[
prl,
twocolumn,
superscriptaddress,
amsmath,
amssymb
]{revtex4-2}

\usepackage[colorlinks=true,
            citecolor=blue,
            linkcolor=blue,
            urlcolor=blue]{hyperref}

\usepackage[dvipsnames]{xcolor}
\usepackage{graphicx}
\usepackage{bm}
\usepackage[mathscr]{eucal}

\usepackage[english]{babel}
\usepackage[normalem]{ulem}
\usepackage{slashed}
\usepackage{color}
\usepackage[colorlinks=true,citecolor=blue,linkcolor=blue,urlcolor=blue]{hyperref}

\usepackage{bibunits}
\defaultbibliographystyle{apsrev4-2}
\defaultbibliography{master_center_vortices_PRL_direct_journal_links_v2}

\newcommand{\Sc}{S_{\mkern-1.5mu\circ}}
\newcommand{\Scm}{S_{\mkern-2.5mu\circ}^{-1}}
\newcommand{\Oc}{{\Omega_{\mkern-1.5mu\circ}}}

\begin{document}

\begin{bibunit}

\title{Dynamical Quarks in the Ensemble of Center Vortices with Monopole Defects: \\ Color Confinement Beyond External Probes}

\author{L. E. Oxman}
\affiliation{
Instituto de F\'isica, Universidade Federal Fluminense, 24210-346 Niter\'oi, RJ, Brasil.}

\date{\today}

\begin{abstract}

We consider the interaction of dynamical quarks with the ensemble of
oriented and nonoriented center vortices proposed to describe
flux-tube formation between quark probes in pure Yang--Mills theory.
The quark sector is described in
terms of bosonized color currents. Within this fluid-like framework, the infrared vortex--monopole condensate restricts finite-energy configurations to be color neutral. Furthermore, neutral distributions of constituent color densities embedded in the condensate generate frustrated regions, providing a mechanism for their localization into finite-size configurations. The resulting picture is consistent with
lattice evidence indicating that center vortices play a central role in
shaping the hadron spectrum.

\end{abstract}

\maketitle

\textit{Introduction}---Understanding confinement in Quantum Chromodynamics (QCD), where quarks are dynamical degrees of freedom, requires explaining why the fundamental quark and gluon fields give rise only to finite-energy, localized, colorless states. This remains one of the central open problems in theoretical physics.

In contrast, in pure Yang--Mills (YM) theory, confinement is characterized by order and
disorder parameters for: the Wilson, Polyakov, and
't~Hooft loops
\cite{Wilson1974,Polyakov1977,tHooft1978}. These observables characterize different aspects of the confining
phase, with the Wilson loop describing the interaction between
external color probes. Their behavior, together with other infrared
properties of the gauge theory, has been extensively investigated in
lattice simulations, establishing the static potential, string
breaking, $N$-ality, and $k$-string tensions, while revealing the center-vortex vacuum structure, as well as infrared field-strength and gluon correlators
\cite{
Bali2001,
Kratochvila2003,
LuciniTeper2001,
LuciniTeperWenger2004,
Debbio1996,
DEliaDiGiacomoMeggiolaro2003,
CucchieriMendes2008}. These developments have been accompanied by infrared descriptions
based on percolating center vortices and on monopole condensation   \cite{
Debbio1997,Forcrand1999,
Langfeld1998,
Engelhardt2000,
Reinhardt2002Topology,
Greensite2020,
DiGiacomo2000,
BoykoPolikarpovZakharov2003,
SakumichiSuganuma2014}. For a detailed recent picture of the geometry of the center-vortex ensemble, see Ref. \cite{Mickley2024}.

\begin{figure}
 \begin{center}
\includegraphics[width=0.23\textwidth]{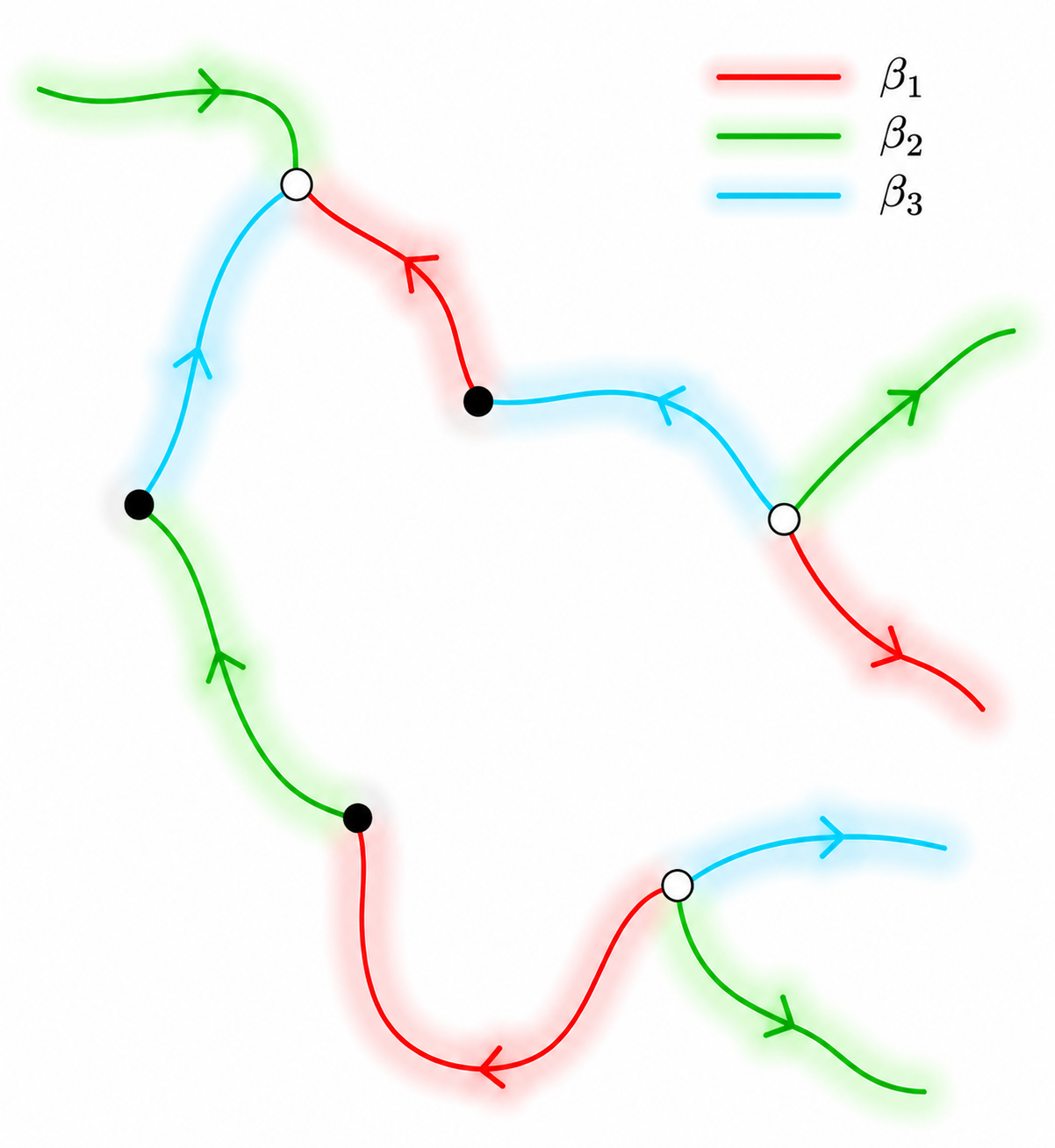}
\end{center}
\caption{For $N=3$, elementary center vortices carry one of the three defining magnetic weights $\beta_i$ ($\beta_1+\beta_2+\beta_3=0$). \\
In 4D, we depict a time slice of the modeled  network.}
\label{land}
\end{figure}

In a series of works
\cite{
Oxman2018,
Junior2020, Junior2021,
Junior:2022WaveFunctional,Oxman:2023Prospecting,JuniorOxman2025Weingarten},
we proposed an infrared framework for pure YM theory based on center
vortices and monopoles forming collimated chains. The Cartan monopole component plays a central role in the formation of finite-width confining flux tubes exhibiting the correct asymptotic properties. Typical correlations incorporated into the ensemble are displayed in
Fig.~\ref{land}. At center monopoles (white dots), three 
vortices carrying center charge $1$ are matched. The  main process considered corresponds to three distinct elementary Cartan fluxes. At Cartan monopoles (black dots), the flux changes orientation within the Cartan subalgebra.
In essence, the Goldstone modes of the center-vortex condensate are gauge fields \cite{Rey1989}, while the monopoles in the chains are described by minimally coupled fields. These are precisely the ingredients required by Derrick scaling \cite{Derrick1964,MantonSutcliffe2004} to stabilize finite-width flux tubes.
 More recently, this approach was shown to account for the infrared mass gap encoded in gauge-invariant field-strength correlators \cite{JuniorKreinOxmanSoares2026PRL}. 

In this Letter, we investigate whether this picture could also explain
confinement in QCD. This possibility is further motivated by lattice evidence
connecting center vortices with the low-lying hadron spectrum
\cite{Trewartha:2017}. 

To address this question, we embed dynamical quarks into the
percolating center-vortex ensemble. The corresponding quark
determinant admits an exact bosonized representation in terms of conserved color currents. This leads to a fluid-like description in
which the currents associated with the defining weights of $SU(N)$ 
interact with the vortex--monopole condensate, providing a framework to investigate how finite-energy localized colorless states emerge.

\textit{QCD and the Yang--Mills ensemble.}---Although the infrared modeling is formulated by parametrizing the simplest elementary processes involving locally Abelian fluxes, it is useful to place the discussion within a first-principles non-Abelian framework.

In Ref.~\cite{OxmanSantosRosa2015},
the obstruction posed by Singer's theorem
\cite{Singer1978}
was proposed as the origin of the Yang--Mills (YM) ensemble. According to this theorem, the space of gauge-field configurations $\{A\}$ in YM theory does not admit a globally defined section. Thus, a global gauge-fixing condition, and hence a global Faddeev--Popov quantization, cannot be defined without encountering Gribov copies
\cite{Gribov1978}.
Singer's work also points to a possible way out: a quantization built from a superposition over local sectors.
Along these lines, in Ref. \cite{OxmanSantosRosa2015}, the YM partition function 
was written in terms of disjoint sectors,  $
Z_{\rm YM}
=
\sum_{\Sc}
Z^{\Sc}_{\rm YM}
$.  
The labels $\Sc$ are $SU(N)$ mappings that contain topological defects. In particular, they can describe distributions of oriented and nonoriented center vortices. For example, an oriented elementary center vortex corresponds to $\Sc(\chi)$, where $\chi$ changes by $2\pi$ along any loop linking $\mathcal W$ (a worldline in 3D or a worldsheet in 4D), and $\Sc(2\pi)=e^{-i\frac{2\pi}{N}}\Sc(0)$. Renormalizability was established in both the vortex-free sector
\cite{FiorentiniEtAl2020}
and the center-vortex sectors
\cite{Fiorentini2022}.

Averages of observables admit an analogous decomposition. In particular, we can take the fermionic partition function
\begin{equation}
Z_{\rm F}[A]
=
e^{-W_{\rm F}}
=
\int D\psi D\bar\psi\,
e^{-S_{\rm F}}
=
\det(\slashed{D}+m)\;,
\end{equation}
to obtain the ensemble representation of the QCD partition function,
\begin{gather}
Z_{\rm QCD}
=
\sum_{\Sc}
Z^{\Sc}_{\rm QCD}\; ,
\quad
Z^{\Sc}_{\rm QCD}
=
\int_{\vartheta_{\Sc}}
[DA]\,
e^{-(S_{\rm YM}+W_{\rm F})}\;.
\nonumber
\end{gather}

\textit{Bosonized quark sector}---Bosonization is a well-established framework that has been developed over more than fifty years. Emerging from the study of low-dimensional condensed-matter systems, it encompasses exact Abelian and non-Abelian formulations in $(1+1)$D as well as descriptions of topological phases in $(2+1)$D. Extensions to $(3+1)$ dimensions have also been proposed. In the higher-dimensional context, the goal is to construct a dual theory in which the correlations of fermionic currents are reproduced by currents that are linear in a bosonizing field \cite{FradkinSchaposnik1994,BurgessQuevedo1994,FoscoSchaposnik1997,LeGuillouMorenoNunezSchaposnik1997,Quevedo1997}.
Although closed expressions for the dual action $S_{\rm B}$ are known only in special cases, the current map can be obtained exactly. We will use this construction in both $(2+1)$D and $(3+1)$D. In the former case, the Dirac fields are taken to be four-component spinors, so as to preserve parity.

The Euclidean partition function in the presence of a non-Abelian $A$ can be rewritten in an exact bosonized form \cite{LeGuillouMorenoNunezSchaposnik1997}. Applying this procedure for $A \in \vartheta_{\Sc}$,
\begin{gather}
Z_{\rm F}[A] 
=  
\int D\Oc\;
e^{-S_{\rm B }[\Oc, \Sc]}
\exp\!\left[\frac{i}{(d-2)!}
\int d x\;
 \bigl(
\Oc , \mathcal{F}(A)
\bigr)
\right] \;,
\nonumber
\end{gather}
where $\mathcal{F}(A)$ denotes the dual field strength.
Here, $\Oc_\mu$ in $(2+1)$D and $\Oc_{\mu\nu}$ in $(3+1)$D are Lie-algebra-valued fields (for details, see Supplemental Material). The bosonized action satisfies
\begin{equation}
S_{\rm B}[U\Oc U^{-1},U\Sc]
=
S_{\rm B}[\Oc,\Sc] \;.
\end{equation}
That is, the physics depends only on the equivalence class represented by $\Sc$. The functional derivative with respect to $A$ leads to identifying the fermionic current $J_\mu(\psi) = J_\mu^a T_a$,  $J_\mu^a = \bar\psi \gamma_\mu T_a \psi $ with the bosonized currents
\begin{equation}
 J_\mu(\Oc) = \begin{cases}
\epsilon_{\mu \nu \rho} D_\nu {\Oc}_\rho 
& \text{in 3D\;,} \\
\frac12
\epsilon_{\mu\nu \rho \sigma }
D_{\nu}{\Oc}_{\rho \sigma }  
& \text{in 4D\;.}
\end{cases} 
\label{d-curr}
\end{equation}
Gauge invariance implies that the expectation values of both $J_\mu(\psi)$ and $J_\mu(\Oc)$ satisfy a covariant conservation law, reflecting the fact that the non-Abelian gauge field carries color.

The partition function can be also written in terms of the gauge invariant field $\Omega$, $\Oc = \Sc \Omega \Scm$, and the dual gauge-invariant field strength $\mathcal{G}(A) = \Scm \mathcal{F}(A) \Sc$:
\begin{gather}
Z_{\rm F}[A] 
=  
\int D\Omega\,
e^{-S_{\rm B }[\Omega]} \,
\exp\!\left[\frac{i}{(d-2)!}
\int d x\;
 \bigl(
\Omega , \mathcal{G}(A)
\bigr)
\right]  \;,
\nonumber \\
\begin{split}
e^{-S_{\rm B }[\Omega]}
=
\int_{\vartheta_{\Sc}} DB\;
\det(\slashed{\partial}+m-i\slashed{B})
\,\Delta[B]\,
\\
\times
\exp\!\left[
-\frac{i}{(d-2)!}
\int d x\;
\bigl(
\Omega , \mathcal{G}(B)
\bigr)
\right] \;, \quad A \in \vartheta_{\Sc}\;,
\end{split} 
\label{Z-rep-gaugeinv}
\end{gather}
Indeed, replacing the second path-integral in the first, the $\Omega$-integral gives the constraint $\Delta[B]\,
\delta \!\left(
\mathcal{F}(B) -
\mathcal{F}(A)
\right)
=
\delta(B_\mu-A_\mu)$, so that integration over $B$ recovers $Z_{\rm F}[A]= \det(\slashed{\partial}+m-i\slashed{A})$. 

For Cartan gauge fields, the Dirac operator is diagonal in color space, with $\psi_i$ coupled through $\omega_i \cdot A$, where $\omega_i$, $i=1,\dots,N$, are the defining weights. The fermionic effective action therefore factorizes into Abelian contributions, implying that $W_{\rm F}$ is obtained from the $U(1)$ effective action by promoting the gauge field to lie in the Cartan subalgebra $\mathfrak h$ and taking traces over the color indices (see the Suppl. Mat.).

We will be interested in the interaction between dynamical quarks and the ensemble of Cartan fluxes. These are collimated configurations which are locally Abelian, characterized by a gauge-invariant field strength $\mathcal G(A) \in \mathfrak{h}$, so that the bosonizing field $\Omega$ is also in $\mathfrak{h}$ and we can take $\Delta[B] \equiv 1$. This ensemble is phenomenologically
motivated by the lattice evidence for the center-vortex vacuum structure
and Abelian/monopole-dominated infrared dynamics 
reviewed in Ref.~\cite{Greensite2020}.

Oriented center-vortex fluxes can be described by $A_\mu \in \mathfrak{h}$. In this case, the dual form of the fermionic current is:
\begin{align}
&J_\mu(\psi) \leftrightarrow J_\mu(\Omega) = 
\begin{cases}
\epsilon_{\mu \nu \rho} \partial_\nu \Omega_\rho & \text{in 3D\;,} \\
\frac{1}{2}\epsilon_{\mu\nu \rho \sigma }
\partial_{\nu}\Omega_{\rho \sigma }  & \text{in 4D\;.}
\label{topo-current}
\end{cases} \\
&J_\mu(\psi) =  \sum_i \omega_i \cdot T\, \bar{\psi}_i
\gamma_\mu \psi_i \label{J-color}\;.
\end{align}
Therefore, the local charge conservation is automatically implemented through a topologically conserved current.  

 \textit{Cartan fluxes}---Elementary center vortices with center charge $k=1$ come in $N$ types, carrying total flux $2\pi\,\beta_i\cdot T$, where $\beta_i = 2N\omega_i$. Configurations formed by $N$ center vortices carrying the different weights $\beta_1,\dots,\beta_N$ matched at a center monopole are also included ($N$-matching). This is possible because $\beta_1+\dots+\beta_N=0$.  
 
 For nonoriented center vortices, the gauge field $A_\mu$ is
non-Abelian and describes monopoles attached to pairs of
center-vortex branches, where the total flux changes from
$2\pi\,\beta_i\cdot T$ to $2\pi\,\beta_j\cdot T$. That is, the
monopole charge is proportional to the root
$\alpha_{ij}=\omega_i-\omega_j$. Such configurations can also have $\mathcal{G}(A)$ mainly in $ \mathfrak h$. In this sector, the topological current in
Eq.~\eqref{topo-current} bosonizes the dressed gauge-invariant
fermionic current $\Scm J_\mu(\psi)\Sc$ (see the Supplemental
Material).
In this case, the Cartan contribution
$S_{\rm B}[\Omega]$, with $\Omega\in\mathfrak h$, is expected to
provide the dominant part of the bosonized action, while corrections
localized at monopoles are absorbed into the effective ensemble
measure. 
The nonoriented component is supported by lattice observations in $SU(2)$ \cite{AmbjornGiedtGreensite2000} and is also consistent with topological analyses on nontrivial manifolds \cite{Hayashi2024}. It is also interesting to note that vortex loops alone in a $Z_N$-lattice realization give a Coulomb phase~\cite{NguyenSulejmanpasicUnsal2025}.

\textit{Quark/center-vortex/monopole system}---For collimat\-ed fluxes localized around a distribution $\{\mathcal W\}$ of center-vortex worldsurfaces in 4D (worldlines in 3D), the field strength essentially vanishes at distances larger than the typical transverse localization scale. For a fermion mass $m$, the current correlations are characterized by a length scale of order $1/m$. This scale is inherited by the bosonized action and is expected to set a typical variation scale for the bosonizing field. Therefore, provided the transverse thickness of the collimated center-vortex fluxes remains smaller than $1/m$, the coupling to $\mathcal{G}(A)$ in Eq. \eqref{Z-rep-gaugeinv} is expected to depend predominantly on the integrated center flux carried by each vortex branch. In this regime, we then consider the approximation $\int dx\,
\bigl(\Omega,\mathcal G\bigr)
\approx
\sum
\int_{\mathcal W}
2\pi\Omega\cdot\beta $,
\begin{gather}
    Z_{\rm QCD} \approx   \int D\Omega\;
e^{-S_{\rm B }[\Omega]} \, \sum_{\{\mathcal W\}}\psi_{\{\mathcal W\}}\, e^{i\sum\int_{\mathcal W} 2\pi \Omega \cdot \beta }  \;.
\label{sumW}
\end{gather}
where we adopted the notation $X \cdot Y = X_q Y_q$.
The weighting function $\psi_{\{\mathcal W\}}$ encodes the properties of elementary center vortices, such as tension and stiffness, and also parametrizes the relative importance of the allowed interactions and correlations. Each configuration $\{\mathcal W\}$ is endowed with a distribution of weight labels, and all branches are summed in the exponent.

\textit{Continuum percolating phase}---The description of the quark/center-vortex/monopole system has parallel realizations in 3D and 4D. In the former case, the center-vortex order parameter is a complex diagonal matrix $\Phi(x)$; $N$-matching and the possibility of nonoriented vortices are represented by local interaction vertices; and the bosonizing field is a vector $\Omega_\mu$. In the condensate, the Goldstone mode is a compact scalar $\Lambda$, $\Phi(x)= \vartheta e^{i \Lambda(x)}$.

In four dimensions, the center-vortex condensate is described by a Goldstone mode $\Lambda_\mu$, while the bosonizing Kalb-Ramond field $\Omega_{\mu\nu}$ enters as a frustration of the center-vortex condensate. Details are given in End Matter (see ``General center-vortex phase in 4D''). This is represented in the continuum by the total Lagrangian $\mathcal L
=\mathcal L_{\Omega}+
\mathcal L_{\Lambda\Omega}+\mathcal L_{\zeta\Lambda}$,
\begin{align}
Z_{\rm QCD}
&\approx
\int D\Omega\,D\Lambda\,D\zeta\;
e^{-\int dx\,\mathcal L}
\;, \nonumber
\\[0.1cm]
\mathcal L_{\Lambda\Omega} 
&=
\frac{\gamma\vartheta^2}{4}
\left(
F_{\mu\nu}(\Lambda)
-
2\pi\,2N\,\Omega_{\mu\nu}
\right)^2
\;, \nonumber
\\
\mathcal L_{\zeta\Lambda}
&=
\sum_\alpha
\vartheta^2\,
\overline{D_\mu\zeta_\alpha}\,
D^\mu\zeta_\alpha
+
\mathcal U(\zeta)
\;, \quad D_\mu
=
\partial_\mu
-
i\,\Lambda_\mu\cdot \alpha \nonumber
\\
\mathcal U(\zeta)
&=
\sum_\alpha
\left(
m^2|\zeta_\alpha|^2
+
\tilde{\lambda}|\zeta_\alpha|^4
\right)
\;. \label{L-sum}
\end{align}
Center-vortex $N$-matching is encoded in the fact that 
$\Lambda_\mu$ is a gauge field in the Cartan subalgebra associated with $U(1)^{N-1} \subset SU(N)$ and not with $U(1)^N$.  The charges of the effective monopole fields are the roots $\alpha$ of $\mathfrak{su}(N)$ ($\zeta_\alpha$ is labeled by a positive root). After rotating to Minkowski spacetime, the theory admits a conserved
total energy. For static magnetic configurations, it takes the form
\begin{gather}
E_{\rm QCD}
\approx
E({\Omega})
+
\int d^3x
\left\{
\frac{\gamma \vartheta^2}{4} G^2_{ij} 
+
\sum_\alpha
\vartheta^2 |D_i\zeta_\alpha|^2
+
\mathcal U(\zeta)
\right\} \nonumber \\
 G_{ij} = F_{ij}(\Lambda)-2\pi 2N \Omega_{ij} \;.
\label{Gij} 
\end{gather}

\textit{Noninteracting fermionic matter}---The bosonized action
$\mathcal L_{\Omega}(J)$ is a generally nonlocal and nonlinear functional
of the conserved current. Through the bosonization map in Eq.~\eqref{topo-current}, together with Eq. \eqref{J-color},
configurations where $\Omega$ approaches $\omega_i \cdot T$ describe fermionic densities $J_0(\Omega)$ carrying color $i$. The color-charge density and the total color charge can be written as
\begin{gather}
J_0 = \nabla \cdot \mathbf b  \;, \quad Q
=
\int_{S^2}
d\mathbf S \cdot \mathbf b \;, \quad b_i = \frac{1}{2} \epsilon_{ijk} \Omega_{jk} \;. \label{div.b}
\end{gather}
A finite energy $E(\Omega)$ only requires $J_0 \to 0$ at infinity, so that static configurations supporting a nonzero total color charge are allowed. 

\textit{Including oriented center vortices}---We now turn on the center-vortex condensate with $N$-matching, so that the system is represented by compact gauge-field Goldstone modes $\Lambda \in \mathfrak{h}$. Because of the term $G_{ij}^2$ in Eq.~\eqref{Gij}, finite energy requires the condition $G_{ij} \to 0$ at infinity,
so that asymptotically
\begin{equation}
\Omega_{ij}
\to
\frac{1}{2\pi 2N}
F_{ij}(\Lambda) \;, \qquad b_i \to \frac{1}{2\pi 2N} \epsilon_{ijk} \partial_j \Lambda_k \;,
\label{f-en}
\end{equation}
which automatically enforces vanishing asymptotic color-charge
density $J_0 = \nabla \cdot \mathbf b \to 0$. 
Just as loop variables constructed from
$U(x,y)=e^{i\Lambda(x,y)}$
are well defined on the lattice, the continuum loop variables
\begin{equation}
U(l)
=
\exp\left(
i\oint_l d\mathbf x\cdot\mathbf\Lambda
\right)
\end{equation}
must also be well defined. The compactness of $\mathbf\Lambda$ is
encoded in the multivaluedness property. To characterize it, consider the family of
loops formed by the parallels on a sphere surrounding a point, from a
small loop around the south pole to one around the north pole. As one
moves through this family, the circulation
$\oint_l d\mathbf x\cdot\mathbf\Lambda$ may change by
$2\pi\,2N\,\alpha_{\rm tot} \cdot T$, where
$\alpha_{\rm tot}$ belongs to the root lattice. This leaves $U(l)$
unchanged because
$
e^{i2\pi\,2N\,\alpha_{\rm tot}\cdot T}
=
I_N $.

Equivalently, $\mathbf\Lambda$ may be described by local patches whose
transition functions satisfy
\[
\mathbf\Lambda^{(1)}
-
\mathbf\Lambda^{(2)}
=
2N\,\alpha_{\rm tot} \cdot T\,\nabla\chi ,
\]
where $\Delta\chi=2\pi$ around closed loops contained in the overlap
of the two patches. Under these conditions, $U(l)$ is independent of
the choice of spanning surface bounded by $l$. Since the
total color charge is determined by the asymptotic flux of
$\mathbf b$, Eq.~\eqref{f-en} implies that the admissible total color
charge is quantized as
\begin{equation}
Q=\alpha_{\rm tot}\cdot T
\;, \quad
\mathbf b(\mathbf x)
\simeq
\alpha_{\rm tot}\cdot T\,
\frac{\hat{\mathbf r}}{4\pi r^2} 
\qquad
(r\to\infty)\;,
\end{equation}
where we have displayed a typical asymptotic configuration in the
corresponding charge sector.
On the other hand, a state associated with a single defining weight
$\omega_i$ is not an admissible asymptotic configuration. This would produce ill-defined compact variables $U(l)$, as $e^{i2\pi\,\beta_i\cdot T} = e^{-i2\pi/N}$.

This is the direct analogue of the $(2+1)$D case, where finite energy similarly forces $\Omega$ to become a pure gauge, $\Omega\sim\frac{1}{2\pi 2N}\nabla \Lambda$. Single-valuedness of $U$ around a point then quantizes the admissible circulation of $\Omega$ on the root lattice, again excluding an isolated charge $\omega_i$. Details are given in End Matter.

\textit{Including the nonoriented component}---We now turn on the nonoriented component. In this case, besides the asymptotic conditions in Eq. \eqref{f-en},  finite energy
requires the
monopole fields $\zeta_\alpha$ to approach their vacuum manifold. When monopoles proliferate ($m^2<0$), the monopole vacua satisfy
\begin{equation}
D_i(\Lambda)\,\zeta_\alpha
=
0 \;,
\qquad
|\zeta_\alpha|^2
=
{\rm const.} \;. \label{m-con}
\end{equation}
The second condition defines a vacuum manifold given by a product of circles. Since its second homotopy group is trivial, the asymptotic phases of the monopole fields are globally defined on $S^2$. The first condition then implies that each projection $\mathbf{\Lambda}\cdot\alpha$, for every positive root $\alpha$, is asymptotically a regular pure gauge. Consequently, the Cartan gauge field is asymptotically gauge equivalent to zero by a regular gauge transformation. Hence, nontrivial gauge patching at spatial infinity is impossible, and the total color charge given by the flux in Eq. \eqref{div.b} must vanish. The analogous $(2+1)$D mechanism is instructive: the instanton sector renders the vacuum manifold discrete, so no continuously changing multivalued phase $\Lambda$---and hence no nontrivial charge---can occur. Details are given in End Matter.

\textit{Bound colorless states}---The next question is whether
meson-like and baryon-like finite-energy localized states can emerge
under the constraint $Q=0$. Let us consider a color-charge density
composed of localized constituents,
\begin{gather}
J_0(\mathbf x)
=
\sum_c
\rho_c(\mathbf x)\,
\omega_c \cdot T
\makebox[.5in]{,}
\int d^3x\,\rho_c(\mathbf x)
=
1\;.
\end{gather}
For mesonic and baryonic configurations, the constituent weights are
$+\omega,-\omega$ and the weights $\omega_1,\omega_2,\omega_3$ ($ \omega_1 + \omega_2 + \omega_3 =0$), respectively. The corresponding field
$\mathbf b$ may be taken as the curl-free solution of
$\nabla \cdot \mathbf b=J_0$. Away from the localized profiles,
\begin{equation}
\mathbf b(\mathbf x)
\simeq
\sum_c
\omega_c \cdot T\,
\frac{\hat{\mathbf r}_c}{4\pi r_c^2}
\;,
\end{equation}
where $\mathbf r_c=\mathbf x-\mathbf x_c$.
When the constituent densities have negligible overlap, energy
minimization locally favors
$G_{ij}\approx0$,
$D_i(\Lambda)\zeta_\alpha\approx0$, and
$|\zeta_\alpha|^2\approx{\rm const.}$
on partial spherical patches surrounding each constituent. These patches, however, cannot be completed into
disjoint closed surfaces while preserving the same conditions.
Indeed, as discussed above, isolated charges $\omega_i$ are excluded,
and the monopole fields $\zeta_\alpha$, with
$\alpha \cdot \omega_i\neq0$, cannot simultaneously remain in the
vacuum manifold and be single valued on a complete surface enclosing
an individual constituent.

The resulting mismatch is therefore expelled into a finite
frustrated region connecting the constituents. Within this region,
$G_{ij}\neq0$, while the corresponding monopole fields are driven
away from the vacuum manifold and vanish along one-dimensional loci.
For a mesonic configuration, this locus is a line joining the two
constituents, whereas for a baryonic configuration it forms a
$Y$-shaped junction. Depending on the parameters, the frustrated
region may be realized as interconnected flux tubes or as a wider
but finite bag-like domain, whose energy scales with its
characteristic size as $\sigma R$ or $BR^3$, respectively. In either
case, the condensate localizes the unavoidable frustration rather
than allowing it to spread throughout space.

The preceding discussion was performed for fixed color-charge densities. The quark dynamics is encoded in the bosonized action $S_{\rm B}[\Omega]$, which can be rewritten as (see Suppl. Mat.)
\[
e^{-S_{\rm B}[\Omega]}
=
\int D\psi D\bar\psi\,
e^{-\int dx\,
\bar\psi
(\slashed{\partial}+m)
\psi}
\,
\delta\!\left(
J(\psi)-J(\Omega)
\right)
\;.
\]
For static configurations persisting for a time $T\to\infty$,
$S_{\rm B}[\Omega]=E(\Omega)T$, where $E(\Omega)$ is the excess
fermionic free energy required to realize the conserved-current
distribution $J_\mu(\Omega)$. For a localized configuration of
characteristic size $R$, dimensional analysis in the light-quark
regime suggests a contribution of order $R^{-1}$, up to logarithmic
corrections. Hence, the fermionic excess free energy tends to oppose
localization.

For baryons, the variational problem must be formulated at fixed baryon number. This constraint can be written in terms of the bosonizing field $\Omega_0$ for the quark-number $U(1)$ current $\sum_i\bar\psi_i\gamma_\mu\psi_i$, with
$\Omega=\Omega_0I+\Omega_qT_q$. The center-vortex condensate couples only to $\Omega_qT_q$, while the exact functional $S_{\rm B}[\Omega]$ couples both sectors.

The central question is whether minimizing over the fields and the
positions $\{\mathbf x_c\}$, with the profiles free to relax, produces
nontrivial finite-size local minima, resulting from the competition
between both energy contributions. If such minima exist, they would
provide the classical configurations underlying finite-size hadronic
states. Otherwise, the energy landscape as a function of $\{\mathbf x_c\}$
exhibits confining behavior at large separations, and the associated
path integral over $\{\mathbf x_c\}$ defines an equivalent Hamiltonian
problem, expected to possess a discrete bound-state spectrum.

\textit{Conclusions}---As is well known, the Higgs and confining
regimes are continuously connected in theories with fundamental
matter~\cite{FradkinShenker}. Correspondingly, the fact that physical asymptotic states are color singlets in both regimes cannot by itself characterize confinement, as emphasized in Ref.~\cite{Greensite}. The relevant infrared question distinguishing the confining and Higgs regimes is therefore whether QCD dynamics restricts finite-energy colorless excitations to remain localized.

In this Letter, we addressed this problem by coupling dynamical quarks
to the infrared center-vortex/monopole ensemble previously proposed
to describe the Yang--Mills vacuum. The emerging picture is fluid-like: bosonized color currents are embedded in the center-vortex/monopole condensate, which acts as the confining medium.

In this setting, we found that it is the finite-energy asymptotic conditions that dynamically restrict configurations to be colorless. Furthermore, when such an excitation is composed of separated colored constituents, the associated frustration is confined to a finite region connecting the constituents, whose energy grows with their separation.
 The same mechanism admits parallel realizations in
$(2+1)$D and $(3+1)$D, with the corresponding bosonizing fields,
Goldstone modes of the center-vortex condensate, and monopole sectors
playing analogous roles.

In this way, the infrared vacuum governed by the mixed center-vortex/monopole condensate provides a common origin for confining flux tubes between external probes, the Yang--Mills mass gap encoded in gauge-invariant field-strength correlators, and the localization of dynamical quarks into finite-size hadronic states. This picture is consistent with lattice evidence that center vortices play a central role in the low-lying hadron spectrum~\cite{Trewartha:2017}.

\newpage

\textit{Acknowledgments}---The author thanks Jeff Greensite and Nick Manton for helpful discussions. Financial support from the Brazilian agency CNPq under Contract No. 309971/2021-7 is gratefully acknowledged.

\putbib

\appendix

\section*{End Matter}

\textit{Bosonization}---This procedure is rooted in the gauge invariance of the fermionic partition 
function $Z_{\rm F}[A]$ in the presence of an external gauge field $A$. This makes it possible to write it as
a path integral over a gauge field $B$, subject to a constraint that
equates $A$ and $B$ up to a gauge transformation. The constraint is
exponentiated through a Lagrange multiplier, the bosonizing field, which
yields the current bosonization map and the path-integral representation of
the bosonized action. In both representations, the full non-Abelian current
is covariantly conserved on average, $D_\nu\langle J^\nu\rangle=0$, as a
consequence of gauge invariance. 

The Cartan sector is special: the current
is simply conserved on average, $\partial_\nu\langle J^\nu\rangle=0$.
Moreover, the bosonized current satisfies $\partial_\mu J^\mu(\Omega)=0$
\emph{identically}---an off-shell identity, independent of any
state or background. This identical conservation extends to locally Abelian sectors: dressing $J_\mu(\psi)$ produces a gauge-invariant current
$\Scm J_\mu(\psi)\Sc$, whose bosonized form is again topologically
conserved (see Suppl.~Mat.). In three and four dimensions this allows a charge
to be defined without reference to the equations of motion or to an
expectation value, as follows,
\[
J_0
=
\epsilon_{ij}\partial_i\Omega_j
\makebox[.9in]{\rm and}
J_0
=
\frac12\epsilon_{ijk}\partial_i\Omega_{jk}\;,
\]
so the corresponding Cartan color charges are given by the circulation and
flux of the bosonizing fields,
\begin{gather}
Q
=
\oint_{S^1_\infty}
dx_i\,\Omega_i\;,
\quad
Q
=
\int_{S^2_\infty}
d\mathbf S \cdot \mathbf b\;,
\quad
b_i
=
\frac12\epsilon_{ijk}\Omega_{jk}\;. \nonumber
\end{gather}
A finite-energy configuration requires the bosonized energy density
$\mathcal E(\Omega)$ to vanish asymptotically, so $J_0(\Omega)\to0$ and the
bosonizing fields become locally exact,
\[
\Omega_i
\to
\partial_i\xi\;,
\qquad
\Omega_{ij}
\to
\partial_i\xi_j-\partial_j\xi_i\;,
\]
respectively. The three-dimensional charge is then determined by the
variation of $\xi$ along $S^1_\infty$, and the four-dimensional charge by
the flux of the corresponding closed two-form through $S^2_\infty$.

\bigskip

\textit{Dynamical quarks and the ensemble in $(2+1)$D}---In the center-vortex/monopole sector, the weight
$\psi_{\{\mathcal W\}}$ 
in Eq.~\eqref{sumW} is modeled from the
tension, stiffness, and contact interactions of the vortex
worldlines~\cite{Lemos2012}. Summing the resulting ensemble for each
defining weight $\omega_i$ admits an effective field-theory
representation~\cite{Junior2020, Junior:2022WaveFunctional} in terms
of $N$ complex scalars $\phi_i$, organized into the diagonal matrix
$\Phi$,
\begin{equation*}
\mathscr{Z}[\Omega]
=
\int D\Phi\;
e^{-\int d^3x\left[{\rm tr}\left((D(\Omega)\Phi)^\dagger
D(\Omega)\Phi\right)+V(\Phi,\Phi^\dagger)\right]}\;,
\end{equation*}
with $D_\mu(\Omega)=\partial_\mu-i2\pi2N\,\Omega_\mu$, and
\begin{equation*}
V
=
\lambda\,{\rm tr}(\Phi^\dagger\Phi-\vartheta^2I_N)^2
-\xi(\det\Phi+{\rm c.c.})
-\nu\,\sum_{i<j}
\bar\phi_j(x)\phi_i(x)\;.
\end{equation*}
The quartic term enforces tension and excluded-volume repulsion,
the determinant term implements $N$-matching, and the last,
monopole-induced term mixes different weights $\omega_i,\omega_j$. As
these interactions are successively turned on, the vacuum manifold is
reduced from $U(1)^N$ to $U(1)^{N-1}$ and finally to the discrete
$Z(N)$ vacua. For static configurations, the
conserved energy is
\begin{equation}
E_{\rm QCD}
=
E(\Omega)
+
\int d^2x\,
\left[
(D_i\Phi)^\dagger D_i\Phi
+
V(\Phi,\Phi^\dagger)
\right]\;.
\label{energy-3d}
\end{equation}

We initially turn on
the oriented center-vortex sector, in the condensed phase ($\vartheta^2>0$). The vacuum manifold
is determined by $\Phi^\dagger\Phi=\vartheta^2 I_N$, so its first
homotopy group is nontrivial and vortex-like configurations are
possible. Finite energy requires $\Phi\to\vartheta U$ at infinity,
where $U$ belongs to $U(1)^N$, together with a vanishing covariant
derivative, $D_i(\Omega)\Phi\to0$. Therefore,
\begin{equation}
(\Omega_x,\Omega_y)
\to
-\frac{i}{2\pi\,2N}
(\nabla U)U^\dagger
=
\frac{1}{2\pi\,2N}
\nabla\Lambda\;,
\end{equation}
which automatically implies
$J_0\to0$ and vanishing asymptotic energy density $\mathcal E(\Omega)$.

When $N$-matching is turned on, $\xi>0$, the vacuum manifold is
further restricted,
\begin{equation}
\Phi^\dagger\Phi
=
\vartheta^2 I_N\;,
\qquad
\det\Phi
=
1\;,
\end{equation}
so the vacua are given by $U \in U(1)^{N-1}$. 

Single-valuedness of $U(\varphi)$ at infinity, where $\varphi$ is the polar angle, constrains the allowed
asymptotic windings and therefore the total color charges. After one turn around spatial infinity, $U(2\pi)=U(0)$, which implies
\[
\Delta\Lambda
=
2\pi\,2N\,\alpha_{\rm tot}\cdot T
\quad\Rightarrow\quad
Q
=
\oint dx_i\,\Omega_i
=
\alpha_{\rm tot}\cdot T\;.
\]
A charge associated with a single fundamental weight $\omega_i$ does
not belong to the adjoint magnetic lattice: if $U(\varphi)$ were
characterized by $\omega_i$, then after one turn around a circle at infinity it would change by a center element, $U(2\pi)=e^{-i2\pi/N}U(0)$, so
$\Phi$ would become discontinuous if it remained everywhere inside the vacua manifold. See \textit{Including oriented center vortices} in the main text for the similar reasoning in $(3+1)$D. 

This obstruction can be avoided only if $\Phi$
leaves the condensate somewhere, for instance if $\Phi\simeq0$ around a
half-line extending from the origin to infinity---such an isolated
configuration therefore has infinite energy.

When the nonoriented interaction is also turned on, $\nu\neq0$, the vacua become discrete, $\Phi\to\vartheta z I_N$, $z\in Z(N)$, so
the continuous Cartan phase is no longer available to accommodate
nontrivial asymptotic windings. Finite energy therefore requires
\begin{equation}
Q
=
\int d^2x\,J_0
=
0\;.
\label{neutrality-3d}
\end{equation}

Fixed color-neutral localized constituent densities may be parametrized by
\begin{equation*}
\Omega
=
\frac{1}{2\pi}
\sum_c
a_c\nabla\chi_c\,
\omega_c\cdot T\;.
\end{equation*}
A fundamental-weight winding around each constituent produces a phase mismatch that the condensate must resolve, in a manner set by
the relative stiffness of the modulus and phase sectors: a soft center-vortex
condensate can locally suppress $\Phi^\dagger\Phi<\vartheta^2I_N$, while a stiff one instead channels the mismatch through the phase $\Lambda$.

When quarks are dynamical, under the Derrick scaling $\Phi(\mathbf x)\to\Phi(\lambda\mathbf x)$,
$\Omega_i\to\lambda\,\Omega_i(\lambda\mathbf x)$, the potential energy
scales as $\lambda^{-2}$ against a scale-invariant gradient term, so shrinking is favored, while the quadratic contribution of the bosonized
energy $E^{(2)}(\Omega) = \int d^2x\, J_0(-\nabla^2)^{-1}J_0 = \int d^2x\, \Omega^2$ (transverse $\Omega$) stays approximately scale invariant as long as the configuration is
much larger than $m^{-1}$. Below this
scale, the determinant becomes effectively massless, and the
quadratic fermionic action is dominated by the universal nonlocal
operator $1/\sqrt{-\nabla^2}$ \cite{BarciFoscoOxman1996},
\[
E^{(2)}(\Omega) \sim \int d^2x\, J_0 \frac{1}{\sqrt{-\nabla^2}}J_0 \;,
\]
giving an energy scaling 
$\lambda\sim R^{-1}$. Together with the nonlinear
terms, this competition can produce a finite-size localized state. In this respect, it is interesting to note that a stabilized bag-like state occurs in a distinct but related $(2+1)$D model formulated in terms of a single complex field $V$ \cite{FoscoKovner2001},
whose center-vortex sector is governed by the vortex-field
interaction $V^N+\bar V^{\,N}$~\cite{tHooft1978} and matter sector by the energy functional $\int d^2x\, J_0^2$.

\bigskip

\textit{General center-vortex phase in $4D$}---The previous description can be understood on the lattice, where localized departures from the center-vortex condensate can also be formulated. In Ref.~\cite{JuniorOxman2025Weingarten}, using the Weingarten representation for the sum over surfaces, the 4D mixed ensemble was represented by
\begin{gather}
\sum_{\{\mathcal W\}}\psi_{\{\mathcal W\}}\,
e^{i\sum\int_{\mathcal W} 2\pi \Omega\cdot\beta}
\to
\int D\Phi D\zeta\,
e^{-\left(W_{\rm v}[\Phi,\Omega]+W_{\rm m}[\zeta,\Phi]\right)}
\;,
\nonumber\\
W_{\rm v}[\Phi,\Omega]
=
-\gamma\sum_p{\rm tr}\,
\Big(e^{-i2\pi 2N \Omega(p)}\Phi(p)\Big) \nonumber \\
+\sum_{\{x,y\} } \left\{  {\rm tr}\, 
\Big(\eta\Phi^\dagger\Phi+\lambda(\Phi^\dagger\Phi)^2\Big)
-\xi
\Big(\det\Phi(x,y)+ \mathrm{c.c.}\Big) \right\} 
\;.
\nonumber 
\end{gather}
Here, \(\Phi(x,y)\) is an \(N\times N\) diagonal matrix, with complex diagonal entries  $\phi_i(x,y)$. When $W_{\rm m}=0$ and $\xi =0$, the modelling reduces to $N$ independent ensembles of closed
center-vortex worldsurfaces, with excluded volume effects, carrying magnetic weights \(\beta_i\). When $\xi \neq 0$, the determinant term implements
center-vortex \(N\)-matching, allowing open surfaces carrying the $N$ different $\beta_i$ to join at common links. The monopole sector $W_{\rm m}(\zeta, \Phi)$ leads to transitions between vortex branches carrying different weights \(\beta_i\) and \(\beta_j\) (nonoriented component). 

When the renormalized center-vortex
tension becomes sufficiently small, a condensate is formed
\begin{gather}
\Phi(x,y)
=
\vartheta U(x,y) \;, \quad 
U(x,y)
=
e^{i\Lambda(x,y)}
\in U(1)^{N-1} \;, \nonumber
\end{gather}
$\Lambda(x,y)  \in \mathfrak h$. If modulus fluctuations are suppressed, only the softer Goldstone
modes survive and the lattice action reduces to ($a$ is the lattice spacing)
\begin{gather}
W_{\rm v}[\Phi,\Omega]+W_{\rm m}[\zeta,\Phi]
\approx
\gamma\vartheta^2
\sum_p{\rm tr}
\Big(I-e^{-i2\pi 2N \Omega(p)}U(p)\Big) \nonumber \\
+\sum_{x,\mu,\alpha}
\vartheta^2
(\Delta_\mu\zeta_\alpha)^\dagger
\Delta_\mu\zeta_\alpha
+ \sum_{x,\alpha}
\Big(
a^2\mu^2\,|\zeta_\alpha|^2
+\tilde\lambda\,|\zeta_\alpha|^4
\Big)
\;,
\nonumber\\
\Delta_\mu\zeta_\alpha
=
e^{i\Lambda(x,x+\mu)\cdot\alpha}
\zeta_\alpha(x+\mu)
-\zeta_\alpha(x)\;.
\end{gather}
Thus, contact is made with $\mathcal L_{\Lambda\Omega}
+ \mathcal L_{\zeta\Lambda} $ in Eq. \eqref{L-sum},  which describes the percolating phase.

\end{bibunit}

\clearpage
\onecolumngrid

\begin{center}
\textbf{\large Supplemental Material}
\end{center}
\bigskip

\begin{bibunit}
\section{QCD and the Yang--Mills ensemble}

In Ref.~\cite{Oxman2015}, a partition of the gauge-field configuration
space $\{A\}$ induced by an equivalence relation was proposed. The main
idea is to initially correlate the gauge field $A$ with an auxiliary tuple
$\zeta=(\zeta_1,\zeta_2,\dots)$ formed by adjoint scalar fields. This
mapping $A\to\zeta(A)$ is obtained as the solution to the classical
equations
\begin{equation}
    \frac{\delta S_{\rm aux}}{\delta\zeta}=0 \;.
\end{equation}
Next, a generalized polar decomposition is introduced in terms of a modulus
tuple $q$ and a phase $S\in SU(N)$,
$\zeta=S q S^{-1}\equiv (S q_1S^{-1},S q_2S^{-1},\dots)$. The modulus satisfies $f(q)=0$, where $f$ is a Lie-algebra-valued
function. Applied to $\zeta(A)$, this defines  a mapping $S(A)$. As the phases
$S(A)$ are not always regular, this leads to a nontrivial partition into equivalence classes
$[\Sc]$, based on the equivalence relation 
$S'\sim S$ if $S'=US$ for some regular $U\in SU(N)$. The labels $\Sc$ are class representatives,
containing different distributions of topological defects such as
oriented and nonoriented center vortices. $A_\mu$ belongs to the
sector $\vartheta_{\Sc}$ if it can be transformed by a regular gauge
transformation so that
\begin{equation}
\zeta(A)=\Sc q(A)\Scm \;.
\label{S0map}
\end{equation}
The conditions for this construction to be well defined were discussed in
Ref.~\cite{FiorentiniEtAl2021}.

For a given gauge field, the solution of the auxiliary equations is
introduced into the path integral through
\begin{gather}
1=\int[D\zeta]\;
\delta\!\left(\zeta-\zeta(A)\right)\;,
\qquad
\delta\!\left(\zeta-\zeta(A)\right)
=
J\,
\delta\!\left(
\frac{\delta S_{\rm aux}}{\delta\zeta}
\right)\;.
\end{gather}
Together with Eq.~\eqref{S0map} and the modulus condition $f(q)=0$, this
filters the contribution from $\vartheta_{\Sc}$,
\begin{gather}
Z_{\rm YM}
=
\sum_{\Sc}Z_{\rm YM}^{\Sc}
\makebox[.5in]{,}
\quad
Z_{\rm YM}^{\Sc}
=
\int_{\vartheta_{\Sc}}
[DA]\,
e^{-S_{\rm YM}}\;.
\label{general-1}
\end{gather}

Averages of observables admit an analogous decomposition. In particular,
\begin{equation}
Z_{\rm F}[A]
=
\int D\bar\psi D\psi\,
e^{-S_{\rm F}}
=
e^{-W[A]}
\makebox[.5in]{,}
\qquad
W[A]
=
-\log\det(\slashed{D}+m)\;,
\end{equation}
which gives the QCD partition function
\begin{equation}
Z_{\rm QCD}
=
\int DA\,
Z_{\rm F}[A]\,
e^{-S_{\rm YM}[A]}\;,
\end{equation}
and the corresponding ensemble representation
\begin{gather}
Z_{\rm QCD}
=
\sum_{\Sc}
Z_{\rm QCD}^{\Sc}
\makebox[.5in]{,}
\quad
Z_{\rm QCD}^{\Sc}
=
\int_{\vartheta_{\Sc}}
[DA]\,
e^{-S_{\rm YM}[A]}\,
Z_{\rm F}[A]\;.
\label{partial}
\end{gather}

\section{Fermionic currents and action in dual form}

The functional non-Abelian bosonization procedure in higher dimensions \cite{FradkinSchaposnik1994,BurgessQuevedo1994,FoscoSchaposnik1997,LeGuillouMorenoNunezSchaposnik1997,Quevedo1997} can be applied in each sector $\vartheta_{\Sc}$. We consider fundamental quarks in $d$-dimensional Euclidean spacetime, with action
\begin{equation}
S_{\rm F}
=
\int dx\, \bar\psi \bigl(\slashed{D}+m\bigr)\psi
\makebox[.5in]{,}
\slashed{D} = \gamma_\mu D_\mu
\makebox[.5in]{,}
D_\mu = \partial_\mu - iA_\mu \;,
\label{fund}
\end{equation}
where the external (Hermitian) source $A_\mu$ takes values in the Lie algebra of a compact group $G$. The gauge group is $G=SU_{\rm c}(N)\times U(1)$. The color and flavor degrees of freedom are arranged into the general fermionic field $\psi$. The Dirac matrices satisfy $\{\gamma_\mu,\gamma_\nu\}=2\delta_{\mu\nu}$ and $\gamma_\mu^\dagger=\gamma_\mu$. In $d=3$, the Dirac fields are taken to be four-component spinors so that parity is preserved and no Chern--Simons term is generated.

We can write
\begin{gather}
\bar\psi\slashed{A}\psi
=
J_\mu^aA_\mu^a
=
(J_\mu,A_\mu)
\makebox[.5in]{,}
J_\mu^a(x)
=
\bar\psi\gamma_\mu T_a\psi
\makebox[.5in]{,}
J_\mu=J_\mu^aT_a\;.
\end{gather}
As the fermionic measure is invariant under local gauge transformations
$\psi\to U\psi$ and
$\bar\psi\to\bar\psi U^{-1}$,
\begin{equation}
Z_{\rm F}[A^U]
=
Z_{\rm F}[A]
\makebox[.5in]{,}
A_\mu^U
=
UA_\mu U^{-1}
+iU\partial_\mu U^{-1}\;.
\end{equation}
Using an infinitesimal transformation $U=e^{i\omega}\approx I+i\omega$, the variation
$\delta A_\mu=D_\mu\omega$,
with
$D_\mu\omega=\partial_\mu\omega-i[A_\mu,\omega]$,
yields
\begin{gather}
0
=
\delta\ln Z_{\rm F}
=
\int dx
\left(
\frac{\delta\ln Z_{\rm F}}{\delta A_\mu(x)},
\delta A_\mu
\right)
=
-\int dx
\left(
D_\mu
\frac{\delta\ln Z_{\rm F}}{\delta A_\mu(x)},
\omega
\right)\;.
\end{gather}
for every $\omega$. This is the covariant conservation of the averaged fermionic current,
\begin{equation}
D_\mu\langle J_\mu\rangle=0
\makebox[.5in]{,}
\qquad
\langle J_\mu\rangle
=
-i\frac{\delta\ln Z_{\rm F}}{\delta A_\mu}\;.
\end{equation}

An auxiliary gauge field $B_\mu$ can be introduced,
\begin{equation}
Z_{\rm F}[A]
=
\int DB_\mu\;
\delta(B_\mu-A_\mu)\,
\det(\slashed{\partial}+m-i\slashed{B})\;.
\end{equation}
Instead of enforcing $B_\mu=A_\mu$, the equality of field strengths can be considered,
\begin{equation}
F_{\mu\nu}[B]
=
F_{\mu\nu}[A]
\makebox[.5in]{,}
F_{\mu\nu}[A]
=
\partial_\mu A_\nu-\partial_\nu A_\mu
-i[A_\mu,A_\nu]\;.
\end{equation}
The constraint is implemented by a Lagrange multiplier,
\begin{align}
Z_{\rm F}[A]
&=
\int_{\vartheta_{\Sc}}
DB\,D\Oc\;
\det(\slashed{\partial}+m-i\slashed{B})\,
\Delta[B]
\nonumber\\
&\quad\times
\exp\!\left[
-i
\int dx\;
\frac{1}{(d-2)!}
\left(
\Oc,
\mathcal F(B)-\mathcal F(A)
\right)
\right]
\makebox[.5in]{,}
A\in\vartheta_{\Sc}\;,
\label{...}
\nonumber\\
\mathcal{F}_{\mu_1\dots\mu_{d-2}}[B]
&=
\frac12
\epsilon_{\mu\nu\mu_1\cdots\mu_{d-2}}
F_{\mu\nu}[B]
\makebox[.5in]{,}
\Delta[B]
=
\left|
\det\left(
\epsilon_{\mu_1\cdots\mu_d}
D_{\mu_1}[B]
\right)
\right|\;.
\end{align}
In three and four dimensions, this amounts to a vector field $\Oc_\mu$ and an antisymmetric tensor $\Oc_{\mu\nu}$, respectively. Integration over $\Oc$ gives
\begin{equation}
\Delta[B]\,
\delta\!\left(
\mathcal F(B)-\mathcal F(A)
\right)
=
\delta(B_\mu-A_\mu)\;,
\end{equation}
while integration over $B$ returns $Z_{\rm F}[A]$. Therefore,
\begin{gather}
Z_{\rm F}[A]
=
\int D\Oc\;
e^{-S_{\rm B}[\Oc,\Sc]}
\exp\!\left[
\frac{i}{(d-2)!}
\int dx\;
(\Oc,\mathcal F(A))
\right]
\makebox[.5in]{,}
A\in\vartheta_{\Sc}\;,
\label{Z-rep}\\
e^{-S_{\rm B}[\Oc,\Sc]}
=
\int_{\vartheta_{\Sc}}
DB\;
\det(\slashed{\partial}+m-i\slashed{B})\,
\Delta[B]\,
\exp\!\left[
-\frac{i}{(d-2)!}
\int dx\;
(\Oc,\mathcal F(B))
\right]\;,
\nonumber\\
(\Oc,\mathcal F)\equiv
\begin{cases}
(\Oc_\mu,\mathcal F_\mu)\;, & \text{in 3D}\;,\\
(\Oc_{\mu\nu},\mathcal F_{\mu\nu})\;, & \text{in 4D}\;.
\end{cases}
\end{gather}

From the functional derivative of $Z_{\rm F}[A]$ with respect to $A$, the fermionic current in bosonized form is
\begin{equation}
J_\mu=
\begin{cases}
\epsilon_{\mu\nu\rho}D_\nu\Oc_\rho\;,
& \text{in 3D}\;,\\
\frac12
\epsilon_{\mu\nu\rho\sigma}
D_\nu\Oc_{\rho\sigma}\;,
& \text{in 4D}\;.
\end{cases}
\label{d-curr}
\end{equation}
Here, $D_\mu=D_\mu(A)$. Its covariant conservation follows from
\begin{equation}
D_\mu J_\mu
=
-i[\mathcal F,\Oc]\;,
\end{equation}
where we used $[D_\mu,D_\nu]X=-i[F_{\mu\nu},X]$. On the other hand, gauge invariance of Eq.~\eqref{Z-rep} implies
$\langle[\mathcal F,\Oc]\rangle=0$.

In the general non-Abelian case,
\begin{gather}
S_{\rm B}[U\Oc U^{-1},U\Sc]
=
S_{\rm B}[\Oc,\Sc]\;,
\end{gather}
that is, the physics only depends on the equivalence class, implying the gauge invariance of $Z_{\rm F}[A]$. Note also that he integrand in Eq.~\eqref{Z-rep} is invariant under
\begin{eqnarray}
\begin{cases}
\Oc_\mu\to\Oc_\mu+D_\mu\chi\;,
& \text{in 3D}\;,\\
\Oc_{\mu\nu}
\to
\Oc_{\mu\nu}
+D_\mu\chi_\nu
-D_\nu\chi_\mu \;,
& \text{in 4D}\;.
\end{cases}
\end{eqnarray}

Then, the QCD partition function takes the form
\begin{gather}
Z_{\rm QCD}
=
\sum_{\Sc}
Z_{\rm QCD}^{\Sc}\;,
\nonumber\\
Z_{\rm QCD}^{\Sc}
=
\int_{\vartheta_{\Sc}}
[DA]\,D\Oc\,
e^{-S_{\rm YM}[A]-S_{\rm B}[\Oc,\Sc]}
\exp\!\left[
\frac{i}{(d-2)!}
\int dx\;
(\Oc,\mathcal F(A))
\right]\;.
\end{gather}

\section*{$\vartheta_{\Sc}$ sectors}

Configurations $A_\mu\in\vartheta_{\Sc}$ can be parametrized in the form \cite{Oxman2013b,Oxman2018}
\begin{gather}
\mathrm{Ad}(A_\mu)
=
R\,\mathrm{Ad}(P_\mu)\,R^{-1}
+iR\,\partial_\mu R^{-1}
\makebox[.5in]{,}
R=\mathrm{Ad}(\Sc) \;.
\label{eq:15}
\end{gather}
The variable $\mathrm{Ad}(\Sc)$ is in the adjoint representation, it is associated with a local frame in the Lie algebra, $n_a=\Sc T_a\Scm = T_b R|_{ba} $, which contains defects but is single-valued around any loop. On the other hand, when a loop links a center-vortex worldsurface, $\Sc$ changes by a center element $e^{\pm i2\pi k/N}$, $k=0,\dots,N-1$.

 To write well-defined expressions in the defining representation, it is convenient to introduce the frame-dependent fields $Z_\mu^a$, $a=1,\dots,N^2-1$, through
\begin{gather}
\mathrm{Ad}(Z_\mu)
=
Z_\mu^a M_a
=
iR^{-1}\partial_\mu R
\makebox[.5in]{,}
M_a=\mathrm{Ad}(T_a)\;.
\end{gather}
Thus, defining $Z_\mu=Z_\mu^a T_a$, we can write 
\begin{gather}
A_\mu
=
\Sc(P_\mu-Z_\mu)\Scm\;, \qquad 
G_{\mu\nu}(A)
=
\Scm F_{\mu\nu}(A)\Sc\;.
\label{eq:18}
\end{gather}
The defects are manifested in the field strengths through commutators of ordinary derivatives, which are nontrivial when applied to singular mappings:
\begin{gather}
F_{\mu\nu}(A)
=
\Sc\bigl(F_{\mu\nu}(P)-F_{\mu\nu}(Z)\bigr)\Scm
\makebox[.35in]{,}
G_{\mu\nu}(A)
=
F_{\mu\nu}(P)-F_{\mu\nu}(Z)
\makebox[.35in]{,}
\mathrm{Ad}\bigl(F_{\mu\nu}(Z)\bigr)
=
iR^{-1}[\partial_\mu,\partial_\nu]R\;.
\end{gather}
The field $P_\mu$ must be such that both $A_\mu$ and its field strength are smooth. For collimated configurations, $P_\mu$ is localized around the vortex guiding centers $\mathcal W$.

Writing $\Oc=\Sc\Omega\Scm$, the fermionic partition function and the bosonized action can be expressed in terms of the gauge-invariant fields $\Omega$ and $\mathcal G$:
\begin{gather}
Z_{\rm F}[A]
=
\int D\Omega\;
e^{-S_{\rm B}[\Omega]}
\exp\!\left[
\frac{i}{(d-2)!}
\int dx\;
\bigl(\Omega,\mathcal G(A)\bigr)
\right]
\makebox[.5in]{,}
\mathcal G(A)
=
\mathcal F(P)-\mathcal F(Z)\;.
\label{Z-rep-frame}
\end{gather}

\subsection*{Bosonization in the Cartan sector}

In a general non-Abelian background, the covariant derivative
$D_\mu=\partial_\mu-iA_\mu$ contains nondiagonal generators that mix different color components. However, when $A_\mu$ belongs to the Cartan subalgebra, the Dirac operator is diagonal in color space, so that
\begin{gather}
(\slashed{D}+m)_{ii}
=
\gamma_\mu
\left(
\partial_\mu-i\,\omega_i\cdot A_\mu
\right)
+m\;, \qquad 
\det(\slashed{D}+m)
=
\prod_{i=1}^N
\det(\slashed{D}_i+m)\;,
\nonumber\\
W_{\rm F}
=
-\ln\det(\slashed{D}+m)
=
-\sum_{i=1}^N
\ln\det(\slashed{D}_i+m)\;.
\label{U-fac}
\end{gather}
Thus, using $\sum_i
\omega_i|_{q_1}\cdots\omega_i|_{q_n}
=
{\rm tr}(T_{q_1}\cdots T_{q_n})$, 
the fermionic effective action is obtained from the $U(1)$ effective action by promoting the gauge field to lie in the Cartan subalgebra $\mathfrak h$ of $\mathfrak{su}(N)$ and taking traces over the color indices. The corresponding bosonized current is
\begin{equation}
J_\mu(\Omega)
=
\begin{cases}
\epsilon_{\mu\nu\rho}\,\partial_\nu\Omega_\rho\;,
& \text{in 3D}\;,\\
\frac12
\epsilon_{\mu\nu\rho\sigma}
\partial_\nu\Omega_{\rho\sigma}\;,
& \text{in 4D}\;.
\label{topo-conserved}
\end{cases}
\end{equation}
where $\Omega=\Omega_qT_q$. Thus, local current conservation is automatically implemented through a topologically conserved current. The theory enjoys the  symmetry
\begin{equation}
\begin{cases}
\Omega_\mu
\to
\Omega_\mu+\partial_\mu\chi\;,
& \text{in 3D,}\\
\Omega_{\mu\nu}
\to
\Omega_{\mu\nu}
+\partial_\mu\chi_\nu
-\partial_\nu\chi_\mu\;,
& \text{in 4D.}
\end{cases}
\label{symms}
\end{equation}
More generally, one may also bosonize the quark number current
$\bar\psi\gamma_\mu\psi
=
\sum_i
\bar\psi_i\gamma_\mu\psi_i
\leftrightarrow
\epsilon\partial\Omega_0 $.

\subsection*{Bosonization in locally Abelian sectors}

The local frame is
\begin{gather}
n_q=\Sc T_q\Sc^{-1} \;,
\qquad
n_\alpha=\Sc T_\alpha\Sc^{-1} \;,
\qquad
n_{\bar\alpha}=\Sc T_{\bar\alpha}\Sc^{-1} \;,
\label{Tqn}
\nonumber\\
T_\alpha=\frac{E_\alpha+E_\alpha^\dagger}{\sqrt2} \;,
\qquad
T_{\bar\alpha}
=
-\frac{i}{\sqrt2}
(E_\alpha-E_\alpha^\dagger) \;,
\qquad
[T_q,E_\alpha]=\alpha|_q\,E_\alpha \;.
\label{Eal}
\end{gather}
We now restrict ourselves to the locally Abelian configurations describing the infrared vortex--monopole ensemble, and consider configurations where monopole defects occur only in the Cartan directions $n_q$.

A single oriented center vortex, supported on $\mathcal W$ (a worldline in 3D or worldsheet in 4D), is generated by
\begin{gather}
\Sc=e^{i\chi\,\beta\cdot T} \;,
\qquad
\beta=2N\omega  \;,
\qquad 
A_\mu=\partial_\mu\chi\,\beta\cdot T  \;.
\label{thin-ori}
\end{gather}
where $\chi$ changes by $2\pi$ around any loop linking $\mathcal W$. The Cartan frame is regular, whereas the off-diagonal directions associated with roots satisfying
$\beta\cdot\alpha=\pm1$
rotate once around $\mathcal W$. General oriented configurations are obtained by joining elementary center vortices carrying the defining magnetic weights
$\beta_i=2N\omega_i$. The simplest branching process is the $N$-matching junction, where one vortex of each type meets, so that the associated total Cartan flux vanishes.

A typical nonoriented configuration is generated by the mapping
\cite{Oxman2013b,Oxman2018}
\begin{equation}
\Sc
=
e^{i\chi\,\beta\cdot T}W  \;,
\qquad
W=e^{i\theta\sqrt N\,T_\alpha}   \;,
\label{SW}
\end{equation}
which interpolates between two defining magnetic weights related by a Weyl reflection,
\begin{gather}
W(\pi)^{-1}
(\beta\cdot T)
W(\pi)
=
\beta'\cdot T \;,
\qquad
\beta'
=
\beta
-
2\frac{\alpha\cdot\beta}{\alpha^2}\alpha  \; .
\label{Weyl-t}
\end{gather}
In all these cases,
\begin{align}
-\mathcal F(Z)
=
2\pi
\sum_j
\beta_j \cdot T
\int_{\mathcal W_j}
d^{d-2}\sigma\,
\delta\!\left(x-\bar x_j(\sigma)\right)\;.
\end{align}

For the locally Abelian configurations considered here,
$\mathcal G(A)=\mathcal F(P)-\mathcal F(Z)$
lies in the Cartan subalgebra. Moreover, when $P_\mu$ is Cartan valued,
\begin{equation}
D_\rho(P)\,\mathcal G
=
\partial_\rho\mathcal G
-i[P_\rho,\mathcal G]
=
\partial_\rho\mathcal G\;.
\end{equation}
Consequently, every gauge-invariant contribution constructed from $\mathcal G$ and covariant derivatives gets simplified. For example, the quadratic contribution becomes
\begin{equation}
W_{\rm F}^{(2)}
=
\frac14
\int dx\,
\bigl(
\mathcal G_{\mu\nu},
\Pi(-\partial^2)\,\mathcal G_{\mu\nu}
\bigr)\;.
\end{equation}
This motivates approximating the bosonized action by its Cartan-sector contribution,
\begin{equation}
Z_{\rm F}[A]
=
\int D\Omega\;
e^{-S_{\rm B}[\Omega]}
\exp\!\left[
\frac{i}{(d-2)!}
\int dx\,
\bigl(
\Omega,\mathcal G(A)
\bigr)
\right]
\makebox[.5in]{,}
\Omega\in\mathfrak h\;.
\label{CartanDual}
\end{equation}

The bosonized current admits a direct interpretation. Since
$A_\mu=\Sc(P_\mu-Z_\mu)\Scm$,
the fermionic interaction can be written as
\begin{equation}
(J_\mu,A_\mu)
=
(\Scm J_\mu\Sc,
P_\mu-Z_\mu)\;,
\end{equation}
 On the other hand, integrating by parts the interaction term in Eq.~\eqref{CartanDual} gives
\begin{equation}
\frac{i}{(d-2)!}
\int dx\,
(\Omega,\mathcal G)
=
i\int dx\,
(P_\mu-Z_\mu,
J_\mu(\Omega))\;,
\end{equation}
where $J_\mu(\Omega)$ is the topologically conserved current defined in Eq.~\eqref{d-curr}. Hence,
\begin{equation}
\Scm J_\mu\Sc
\longleftrightarrow
J_\mu(\Omega)\;,
\end{equation}
bosonization identifies the gauge-invariant dressed fermionic current with the topologically conserved current in Eq. \eqref{topo-conserved}.

\section{Center-vortex/monopole sector}

Thick center vortices are represented by worldlines in three Euclidean dimensions and worldsheets in four Euclidean dimensions. The corresponding ensemble average is
\begin{gather}
\mathscr Z[\Omega]
=
\sum_{\{\mathcal W\}}
\psi_{\{\mathcal W\}}
\exp\!\left(
i
\sum
\int_{\mathcal W}
2\pi\,\Omega\cdot\beta
\right)\;.
\label{Zwave}
\end{gather}

\subsection{3D}

Following Ref.~\cite{Lemos2012}, the statistical weights $\psi_{\{\mathcal W\}}$ are modeled in terms of the tension, stiffness and contact interactions of the vortex worldlines. The basic building block is the sum over all worldlines carrying a fixed defining weight $\omega$, with prescribed endpoints, orientations and length $L$,
\begin{align}
\sum_\gamma
e^{-\int_0^L ds\,
\left(
\frac{1}{2\kappa}\dot u^2(s)
+\mu
-i\sigma(x)
-i\,u(s)\,
2\pi\Omega(x(s))\cdot\beta
\right)}
\makebox[.5in]{,}
\qquad
u(s)=\frac{dx(s)}{ds}\;.
\end{align}
The auxiliary scalar field $\sigma(x)$, supplemented with a Gaussian weight, encodes repulsive contact interactions between vortex lines carrying the same flux. This building block satisfies a Fokker--Planck diffusion equation whose small-stiffness solution is
\begin{align}
\langle x|
e^{-LO_\beta}
|x_0\rangle
\makebox[.5in]{,}
\qquad
O_\beta
=
-\frac{1}{3\kappa}
(\partial-i2\pi\Omega \cdot \beta)^2
+\mu
-i\sigma(x)\;.
\end{align}
Integration over the lengths $L$ gives the propagator
$O_\beta^{-1}(x,x_0)$, 
so that the sum over vortex networks admits an effective field-theory representation. In the loop sector, if different weights are uncorrelated and noninteracting,
\begin{align}
\mathscr Z_{\rm loops}[\Omega]
=
\prod_{i=1}^N
\det O_{\beta_i}\;.
\label{loops-en}
\end{align}

This determinant can be represented by introducing $N$ complex scalar fields $\phi_i$, organized into the diagonal matrix $\Phi$. The ensemble then becomes \cite{Junior2020, Junior:2022WaveFunctional}
\begin{gather}
\mathscr Z[\Omega]
=
\int D\Phi\;
\exp\!\left[
-\int d^3x
\left(
{\rm tr}
\bigl(
(D(\Omega)\Phi)^\dagger
D(\Omega)\Phi
\bigr)
+
V(\Phi,\Phi^\dagger)
\right)
\right]\;, \qquad 
D_\mu(\Omega)
=
\partial_\mu
-i2\pi2N\,\Omega_\mu\;.
\label{Z3d}
\end{gather}
Indeed,
\begin{gather}
{\rm tr}
\bigl(
(D(\Omega)\Phi)^\dagger
D(\Omega)\Phi
\bigr)
=
\sum_i
\overline{D(\Omega)\Phi}_{ii}
D(\Omega)\Phi|_{ii}\;,
\\
D(\Omega)\Phi|_{ii}
=
\partial_\mu\phi_i
-i2\pi2N\Omega_{ii}\phi_i
=
\partial_\mu\phi_i
-i2\pi(\Omega \cdot \beta_i)\phi_i\;.
\end{gather}
Therefore, when $V=0$, integrating over $\Phi$ reproduces Eq.~\eqref{loops-en}.

The potential is written as
\begin{gather}
V=V_{\rm v}+V_{\rm m} \;, \qquad V_{\rm v}
=
\lambda
{\rm tr}
(\Phi^\dagger\Phi-\vartheta^2I_N)^2
-
\xi
(\det\Phi+\mathrm{c.c.})\;,
\qquad
V_{\rm m}
=
-
\nu\,
{\rm tr}
(\Phi^\dagger E_\alpha\Phi E_\alpha^\dagger)\;.
\label{wavef}
\end{gather}
The quartic term incorporates the center-vortex tension and excluded-volume interactions, the determinant term implements the $N$-matching property, while $V_{\rm m}$ couples different center vortices through monopole-like instantons. For the root $\alpha_{ij}=\omega_i-\omega_j$, since $E_\alpha$ is only nonvanishing at the entry $ij$,
\begin{gather}
V_{\rm m}
\propto
\sum_{ij}
\bar\phi_j(x)\phi_i(x)\;,
\label{v-in}
\end{gather}
which induces transitions between center-vortex weights $\omega_i$ and $\omega_j$ at an instanton.

Consequently, as the interactions are successively turned on, the vacuum manifold is reduced from
$U(1)^N$
to
$U(1)^{N-1}$,
and finally to the discrete $Z(N)$ vacua.

\subsection{4D}

The four-dimensional construction parallels the three-dimensional
worldline ensemble, the elementary objects now being center-vortex
worldsheets. A convenient lattice regularization is provided by the
Weingarten representation \cite{Weingarten1980}, where closed surfaces are
generated by complex link variables. For noninteracting surfaces with
bare tension $\mu_0$,
\begin{gather}
Z_0
=
\sum_{\mathcal S}
e^{-\mu_0A(\mathcal S)}
\makebox[.5in]{,}
A(\mathcal S)=a^2F
\;,
\end{gather}
which admits the Gaussian representation
\begin{gather}
Z_0
=
\int D\phi\,
\exp\!\left(
\gamma\sum_p\phi(p)
-
\sum_{\{x,y\}}
\phi^\dagger(x,y)\phi(x,y)
\right)
\makebox[.5in]{,}
\ln\gamma^{-1}
=
\mu_0a^2
\;.
\label{wein}
\end{gather}

Following Ref.~\cite{JuniorOxman2025Weingarten}, the Abelian-projected ensemble is
obtained by promoting the complex link variable to the diagonal matrix $\Phi(x,y)$ 
whose entries $\phi_i(x,y)$ generate elementary center-vortex branches carrying the
defining magnetic weights
$\beta_i=2N\omega_i$.
The coupling to the Kalb--Ramond field is introduced by frustrating
each plaquette,
\begin{gather}
\mathscr Z[\Omega]
=
\int D\Phi\,
e^{-W_{\rm v}[\Phi,\Omega]-W_{\rm m}[\zeta,\Phi]}
\;, \nonumber \\
W_{\rm v}
=
-\gamma
\sum_p
{\rm tr}
\!\left(
e^{-i2\pi2N\Omega(p)}
\Phi(p)
\right)
+
\sum_{\{x,y\}}
{\rm tr}
\!\left(
\eta\Phi^\dagger\Phi
+
\lambda(\Phi^\dagger\Phi)^2
\right)
-\xi
\sum_{\{x,y\}}
\left(
\det\Phi(x,y)
+
\mathrm{c.c.}
\right)
\;.
\label{Wv}
\end{gather}
Since $2N\Omega_{ii}
=
\Omega\cdot\beta_i$, $W_{\rm v}$ is the four-dimensional counterpart of the kinetic and
potential terms in Eqs.~\eqref{Z3d} and \eqref{wavef}.

The nonoriented component is incorporated through
\begin{gather}
W_{\rm m}[\zeta,\Phi]
=
-\sum_{l,\alpha}
\bar\zeta_\alpha(x)H_\alpha(x,y)\zeta_\alpha(y)
+\sum_{x,\alpha}
\left(
\tilde\eta\,|\zeta_\alpha(x)|^2
+\tilde\lambda\,|\zeta_\alpha(x)|^4
\right)
\;,
\qquad H_\alpha(x,y)
=
\bar\phi_j(x,y)\phi_i(x,y)
\;,
\end{gather}
which is the four-dimensional analogue of the vertex interaction in
Eq.~\eqref{v-in}. Expanding in powers of $H_\alpha$ generates the
gluing of worldsheet boundaries along monopole worldlines.

When the renormalized surface tension becomes sufficiently small, a
condensate is formed,
\begin{gather}
\Phi(x,y)
=
\vartheta U(x,y)
\makebox[.5in]{,}
U(x,y)
=
e^{i\Lambda(x,y)}
\in U(1)^{N-1}
\;,
\end{gather}
whose naive continuum limit reproduces the effective continuum theory
presented in the main text.

\section{Bosonized energy functional}

The dual functional $S_{\rm B}[\Omega]$ plays a central role in the
construction presented in the main text. When $\Omega$ is restricted to
the Cartan sector $\mathfrak h$, it admits a natural interpretation in
terms of a constrained fermionic partition function. Starting from
\begin{gather}
e^{-S_{\rm B}[\Omega]}
=
\int DB\;
\det(\slashed{\partial}+m-i\slashed{B})\,
\exp\!\left[
-\frac{i}{(d-2)!}
\int dx\,
\bigl(
\Omega,\mathcal F(B)
\bigr)
\right]\;.
\label{SDdef} \\
\det(\slashed{\partial}+m-i\slashed B)
=
\int D\bar\psi D\psi\,
\exp\!\left[
-\int dx\,
\bar\psi
(\slashed{\partial}+m-i\slashed B)
\psi
\right]\;. \nonumber \\
\frac{1}{(d-2)!}
\int dx\,
\bigl(
\Omega,\mathcal F(B)
\bigr)
=
\int dx\,
\bigl(
B_\mu,
J_\mu(\Omega)
\bigr)\;,
\end{gather}
integration over $B_\mu$ imposes the constraint
$\bar\psi\gamma_\mu\psi=J_\mu(\Omega)$. Equivalently,
\begin{equation}
e^{-S_{\rm B}[\Omega]}
\propto
\int D\bar\psi D\psi\,
\delta\!\left[
\bar\psi\gamma_\mu\psi
-
J_\mu(\Omega)
\right]\,
e^{-S_{\rm F}[\bar\psi,\psi]}\;.
\label{constrained}
\end{equation}
Therefore, for a static configuration persisting for (an infinite)
Euclidean time $T$,
\[
S_{\rm B}[\Omega]=E(\Omega)\,T,
\]
where $E(\Omega)$ is the excess fermionic free energy required to
enforce the conserved current distribution $J_\mu(\Omega)$. Any
nontrivial current distribution is expected to have higher free energy
than the unconstrained vacuum. Thus, the minimum is expected at
$J_\mu=0$, and throughout this work we choose the reference point
$E(0)=0$.

For a localized configuration of characteristic size $R$, the massless
theory contains no intrinsic length scale apart from those introduced
by renormalization. Dimensional analysis therefore suggests
\begin{equation}
E(\Omega)
\sim
\frac{C}{R}
\;,
\end{equation}
possibly up to logarithmic corrections. This conclusion does not rely
on any particular approximation to the fermionic determinant.

In the present approach, only the center-vortex/monopole sector is
formulated on the lattice, from which the corresponding infrared
effective description is inferred. The bosonized quark sector is
introduced directly in the continuum and coupled to this effective
ensemble, so the usual lattice sign problem associated with Monte Carlo
sampling of dynamical fermion determinants does not arise. 
In the continuum, the Euclidean massless Dirac operator is
anti-Hermitian, $\slashed{D}^{\,\dagger}=-\slashed{D}$, so its
eigenvalues are purely imaginary. Since
$\{\gamma_5,\slashed{D}\}=0$, the nonzero eigenvalues occur in pairs
$\pm i\lambda$, implying that the eigenvalues of the massive operator
$\slashed{D}+m$ are $m\pm i\lambda$. Hence,
\[
\det(\slashed{D}+m)
=
\prod_{\lambda>0}(m^2+\lambda^2)
\ge0,
\]
so the constrained fermionic determinant defining
$S_{\rm B}[\Omega]$ is real and non-negative.

\subsection{Realizations of $E(\Omega)$}

Concrete realizations of the energy functional satisfying the above
qualitative properties are obtained from
\begin{equation}
E(\Omega)
=
-
\int d^{d-1}x
\left(
J^iK_i+\mathcal L_\Omega
\right)
\:, \qquad
K_\mu
=
-
\frac{\partial\mathcal L_\Omega}{\partial J^\mu}
\;,
\label{EOmega}
\end{equation}
after passing to Minkowski space, where the dual sector is described by
the Lorentz-invariant Lagrangian $\mathcal L_\Omega(J)$.

The quadratic part of the dual action follows directly from the exact
current-current response of the fermionic theory. Since bosonization
reproduces all current correlation functions, the quadratic kernel is
the inverse of the connected current correlator in the transverse
sector. In Euclidean spacetime,
\begin{equation}
S_{{\rm B}}^{(2)}
=
\frac12
\int dx\,
\left(
J_\mu,
\frac{1}
{(-\partial^2)\Pi(-\partial^2)}
J_\mu
\right)\;,
\label{SDquad}
\end{equation}
where $\Pi$ denotes the polarization operator induced by the fermionic
determinant. Under Wick rotation,
$-\partial^2\rightarrow\Box$, Eq.~\eqref{EOmega} yields
\begin{align}
E^{(2)}(\Omega)
&=
\frac12
\int d^{d-1}x
\Bigg[
\left(
J_0,
\frac{1}
{\Box\Pi(\Box)}
J_0
\right)
+
\left(
J_i,
\frac{1}
{\Box\Pi(\Box)}
J_i
\right)
\Bigg]
\;.
\label{Equad}
\end{align}
For static configurations, one simply replaces
$\Box\rightarrow-\nabla^2$.

The explicit form of $\Pi$ depends on the spacetime dimension and the
fermionic content. For fundamental quarks,
$\Pi(k^2)=f(k^2)/(2N)$,
where $f$ is the polarization function associated with a single
$U(1)$ component. The factor $1/(2N)$ follows from Eq. \eqref{weights}.

\paragraph*{Three dimensions.}

We consider a parity-preserving fermionic content with four-component
spinors and opposite parity masses, so that no Chern--Simons term is
induced. In this case,
\begin{equation}
f(k^2)
=
\frac{1}{\pi}
\int_0^1d\alpha\,
\frac{\alpha(1-\alpha)}
{\sqrt{m^2+\alpha(1-\alpha)k^2}}
\;.
\label{Pi3-exact}
\end{equation}
Since $f(k^2)>0$, the corresponding quadratic energy is positive
definite.

\paragraph*{Four dimensions.}

For a Dirac fermion,
\begin{equation}
f(k^2)
=
\frac{1}{2\pi^2}
\int_0^1d\alpha\,
\alpha(1-\alpha)
\ln
\left[
\frac{m^2+\alpha(1-\alpha)k^2}{\mu^2}
\right]
\;.
\label{Pi4-exact}
\end{equation}
The subtraction prescription is chosen so that the renormalized
quadratic kernel remains positive in the momentum range under
consideration.
More generally,
\begin{equation}
S_{\rm B}[J]
=
S_{\rm B}^{(2)}[J]
+
S_{\rm B}^{(4)}[J]
+\cdots,
\end{equation}
where the higher-order contributions, generated by the connected
fermionic current correlators, are nonlinear and
nonlocal. For slowly varying currents they admit a derivative
expansion, while for momenta comparable to the fermion mass the full
nonlocal momentum dependence is expected to become important.

The positivity of the quadratic kernel shows explicitly that the
constrained functional increases for sufficiently small departures from
the vacuum. Away from
the vacuum, the nonlinear nonlocal terms are expected to play an
essential role. In particular, they contain the first interactions
between the Cartan color currents and the $U(1)$ baryonic current, and are therefore expected to
be important in future studies of localized configurations.

\subsubsection*{Colorless configurations}

The neutrality condition does not exclude localized constituent color
densities; it only requires their total charge to vanish. In both
dimensions, we consider distributions of the form
\begin{equation}
J_0(\mathbf x)
=
\sum_c
\rho_c(\mathbf x)\,
\omega_c\cdot T
\makebox[.5in]{,}
\int d^{d-1}x\,
\rho_c(\mathbf x)
=
1
\makebox[.5in]{,}
Q
=
\sum_c\omega_c
=
0\;.
\label{localized-density}
\end{equation}
The mesonic and baryonic choices correspond, respectively, to
$+\omega,-\omega$ and, for $SU(3)$,
$\omega_1,\omega_2,\omega_3$. Near an individual constituent, the
density is approximately aligned with a defining weight, although such
a charge cannot extend to spatial infinity.

\vspace{.05in}
\paragraph*{(2+1)D.}

A local constituent density of the form in
Eq.~\eqref{localized-density} may be generated by 
\begin{equation}
\Omega
=
\frac{1}{2\pi}
\sum_c
a_c\nabla\chi_c\,
\omega_c\cdot T\;.
\end{equation}
The profiles $a_c$ regularize the constituent regions, while the
angular variables $\chi_c$ determine their locations.
As summarized in End Matter, the resulting mismatch must be resolved by either the modulus or the phase sector, depending on their relative stiffness.
If the center-vortex condensate is relatively soft, its modulus may be
suppressed within a finite region. 
The associated gradient and condensation-energy costs depend on the
size of this region and may, in principle, compete with the fermionic
energy. Thus, excitation of the modulus sector can provide a
size-dependent contribution even without the nonoriented interaction.

If the condensate is stiff, however, leaving the condensed manifold is
energetically costly. In the purely oriented phase, the mismatch may
then spread through the continuous Cartan vacuum manifold, and the
reduced Goldstone description contains no evident mechanism forcing it
into a localized finite-energy structure. The nonoriented interaction
changes this situation by lifting the continuous degeneracy and
selecting discrete vacua. If the phase-locking sector is softer than
the modulus sector, it becomes natural to preserve
\begin{equation}
\Phi^\dagger\Phi
\simeq
\vartheta^2 I_N
\end{equation}
and accommodate the mismatch predominantly through the phase
$\Lambda$ of $\Phi$. In particular, around an individual constituent,
$\Lambda$ may satisfy $\Delta\Lambda=0$, with its gradient closely
tracking $\Omega$ over most of the circuit. Across a narrow region it
instead departs from $\Omega$ and interpolates between different
discrete vacua, producing a domain wall in the mesonic case (cf.\ the
$N=2$ realization in Ref.~\cite{FoscoKovner2001}) or a Y-shaped
junction formed by three domain walls in the baryonic case. Depending
on the parameters, the departure from the discrete vacua may instead
extend over a broader bag-like region.

\vspace{.05in}
\paragraph*{(3+1)D.}

The field associated with the localized density in
Eq.~\eqref{localized-density} can be obtained from
\begin{equation}
\mathbf b(\mathbf x)
=
\frac{1}{4\pi}
\sum_c
a_c\,
\omega_c\cdot T\,
\frac{\widehat{\mathbf r}_c}{r_c^2}
\makebox[.5in]{,}
J_0
=
\nabla\cdot\mathbf b\;.
\label{b-localized}
\end{equation}
Although the total charge vanishes, the flux near each constituent is
approximately aligned with a defining weight.

As in three dimensions, the manner in which this frustration is
accommodated depends on the relative stiffness of the different
sectors. However, the soft center-vortex condensate cannot be naturally described in the continuum. This situation could be formulated on the lattice, where the link variables
$\Phi(x,y)$ explore the full complex field space rather than remaining
restricted to their Goldstone components. 

If the center-vortex condensate is stiff, exciting its modulus becomes
energetically costly. In the purely oriented phase, the mismatch may
instead spread through the continuous Cartan Goldstone sector
$\Lambda_i$, and the reduced continuum description contains no evident
mechanism forcing it into a localized finite-energy configuration.
When the monopole sector is turned on, it becomes natural to keep the
vortex condensate everywhere and accommodate the
frustration through the Goldstone and monopole fields.

\subsubsection*{Energy competition and Derrick scaling}

The Derrick scaling already reveals the competition between the different energy contributions.

\vspace{.05in}
\paragraph*{(2+1)D.}

The continuum energy is
\begin{equation}
E_{\rm QCD}
=
E(\Omega)
+
\int d^2x\,
\left[
(D_i\Phi)^\dagger D_i\Phi
+
V(\Phi,\Phi^\dagger)
\right]\;.
\end{equation}
Under the scaling
\begin{equation}
\Phi(\mathbf x)
\rightarrow
\Phi(\lambda\mathbf x)\;,
\qquad
\Omega_i(\mathbf x)
\rightarrow
\lambda\,
\Omega_i(\lambda\mathbf x)\;,
\end{equation}
the center-vortex contributions behave as
\begin{align}
\int d^2x\,
(D_i\Phi)^\dagger D_i\Phi
&\sim
\lambda^0\;,
\qquad
\int d^2x\,
V(\Phi,\Phi^\dagger)
\sim
\lambda^{-2}\;.
\end{align}
Thus, the potential energy favors shrinking of the frustrated region,
while the scalar-gradient energy is scale invariant.

From Eq. \eqref{Pi3-exact}, the quadratic fermionic
contribution is approximately scale invariant in the heavy-fermion
regime. As the characteristic size decreases to
$R\sim m^{-1}$, it crosses over to the nonlocal massless behavior $E_\Omega^{(2)}
\sim
\lambda$, providing an effective pressure against further localization. Also, for large $m$, if nonlinear terms become relevant, the effective energy contains schematically ($\Omega$ transverse)
\begin{equation}
E(\Omega)
\sim
\int d^2x\,
\left(
\Omega^2+\Omega^4+\cdots
\right)\;.
\end{equation}
Since \(
\int d^2x\,\Omega^4\sim\lambda^2,
\)
nonlinearities also oppose collapse. 

\vspace{.1in}
\paragraph*{(3+1)D.}

The continuum energy is
\begin{align}
E_{\rm QCD}
=
E(\Omega)
+
\int d^3x
\Big[
&
\frac{\gamma\vartheta^2}{4}
G_{ij}G_{ij}
+
\vartheta^2
(D_i\zeta_\alpha)^\dagger
D_i\zeta_\alpha
+
\mathcal U(\zeta)
\Big]\;.
\end{align}
The natural Derrick scaling is
\begin{equation}
\Lambda_i(\mathbf x)
\rightarrow
\lambda\,
\Lambda_i(\lambda\mathbf x)\;,
\qquad
\zeta_\alpha(\mathbf x)
\rightarrow
\zeta_\alpha(\lambda\mathbf x)\;,
\qquad
\Omega_{ij}(\mathbf x)
\rightarrow
\lambda^2
\Omega_{ij}(\lambda\mathbf x)\;.
\end{equation}
Accordingly,
\begin{align}
\int d^3x\,
G_{ij}G_{ij}
&\sim
\lambda\;,
\qquad
\int d^3x\,
(D_i\zeta_\alpha)^\dagger
D_i\zeta_\alpha
\sim
\lambda^{-1}\;,
\qquad
\int d^3x\,
\mathcal U(\zeta)
\sim
\lambda^{-3}\;.
\end{align}
The quadratic fermionic
contribution scales as
$E_\Omega^{(2)}
\sim
\lambda$, with logarithmic corrections in the massless regime. 

Therefore the various contributions favor opposite tendencies. The
vacuum energy associated with the frustrated region drives shrinking,
whereas sufficiently light fermions and nonlinear corrections resist
localization. Their competition may generate a preferred finite size.

\section*{Defining and adjoint weights of $\mathfrak{su}(N)$}

The $N$ weights $\omega_i$, $i=1,\dots,N$, of the defining representation are given by the $(N-1)$-tuples
\begin{equation}
\omega_i = (T_1|_{ii}, T_2|_{ii},\dots, T_{N-1}|_{ii}) \;.
\label{wfun}
\end{equation}
That is, the tuple $\omega_i$ is formed by the common eigenvalues of the defining weight vector $|\omega_i \rangle$, which is the column vector with a $1$ at the $i$-th row and zero otherwise:
\begin{gather}
    T_q |\omega_i \rangle = \omega_i |_q |\omega_i \rangle \;.
\end{gather}
As the generators are traceless,
$\omega_1+\dots+\omega_N=0 $.
They also satisfy
\begin{gather}
\omega_i\cdot \omega_j = \frac{N\delta_{ij}-1}{2N^2}
\makebox[.5in]{,}
\sum_{i=1}^N \omega_i|_q\, \omega_i|_p =\frac{\delta_{qp}}{2N}\;.
\label{weights}
\end{gather}
The $N(N-1)$ roots of $\mathfrak{su}(N)$ are
$\alpha_{ij}=\omega_i -\omega_j
$, $ i \neq j$ and satisfy (no sum over $i$)
\begin{equation}
\alpha^2= \alpha \cdot \alpha=1/N \makebox[.5in]{,}
2N \omega_i \cdot \alpha_{ij} =1
\makebox[.7in]{,}
2N \omega_i \cdot \alpha_{jk} = 0~~(j, k \neq i)\;.
\label{rela}
\end{equation}
The positive roots are characterized by $i<j$.

\putbib
\end{bibunit}

\end{document}